\documentclass[cits]{PoS}
\usepackage[]{fontenc}
\usepackage{graphicx}

\makeatletter

\providecommand{\tabularnewline}{\\}
\def\apj{Astrophysical Journal}                 % Astrophysical Journal
\def\apjs{Astrophysical Journal Supplement Series}               % Astrophysical Journal, Supplement
\def\mnras{Monthly Notices of the RAS}             % Monthly Notices of the RAS
\title{Reaction Rate Uncertainties: NeNa and MgAl in AGB~Stars}

\ShortTitle{Reaction Rates and NeNa/MgAl}

\author{\speaker{Robert G. Izzard}\\
        University of Utrecht\\
        E-mail: \email{R.G.Izzard@phys.uu.nl}}

\author{Maria Lugaro\\
        University of Utrecht\\
        E-mail:\email{M.Lugaro@phys.uu.nl}}

\author{Christian Iliadis\\
        University of North Carolina\\
        E-mail: \email{iliadis@unc.edu}}

\author{Amanda Karakas\\
        McMaster University\\
        E-mail: \email{karakas@physics.mcmaster.ca}}

\abstract{We study the effect of uncertainties in the proton-capture reaction rates of the NeNa and MgAl chains on nucleosynthesis due to the operation  of hot bottom burning (HBB) in intermediate-mass asymptotic giant branch (AGB) stars. HBB nucleosynthesis is associated with the production of sodium, radioactive $^{26}\mathrm{Al}$ and the heavy magnesium isotopes, and it is possibly responsible for the O, Na, Mg and Al abundance anomalies observed in globular cluster stars. \\
We model HBB with an analytic code based on full stellar evolution models so we can quickly cover a large parameter space. The reaction rates are varied first individually, then all together. This creates a knock-on effect, where an increase of one reaction rate affects production of an isotope further down the reaction chain. We find the yields of $^{22}\mathrm{Ne}$, $^{23}\mathrm{Na}$ and $^{26}\mathrm{Al}$ to be the most susceptible to current nuclear reaction rate uncertainties.}

\FullConference{International Symposium on Nuclear Astrophysics --- Nuclei in the Cosmos --- IX\\
   June 25-30 2006\\
   CERN, Geneva, Switzerland}

\makeatother
\begin{document}

\section{Introduction}

The Asymptotic Giant Branch (AGB) is the final evolutionary phase
of stars with a mass less than about $8\mathrm{\, M_{\odot}}$. The
core of an AGB star has exhausted its supply of hydrogen and helium
by nuclear burning and is made of carbon, oxygen, neon and perhaps
magnesium, the main helium-burning products. A thin helium layer separates
it from the convective stellar envelope which is mostly hydrogen.
Hydrogen burning in a shell at the base of the envelope dominates
the stellar luminosity, which is many thousands of times that of the
Sun, but is punctuated by thermal pulses \emph{}due to ignition of
the helium layer. At each pulse the third dredge-up \emph{}may occur,
in which helium-burnt material is mixed into the stellar envelope,
polluting it with helium, carbon, $^{22}\mbox{Ne}$ and heavy $s$-process
elements. The lifetime of an AGB star is dominated by rapid mass-loss
which limits it to about a million years, after which the stellar
envelope is ejected into the interstellar medium and a CO white dwarf
remains.

In AGB stars more massive than about $4\mathrm{\, M_{\odot}}$ the
hydrogen burning shell, at a temperature of 60 to 100 million K, extends
into the convective envelope, a process called \emph{hot-bottom burning}
(HBB). The envelope composition is directly affected by the various
hydrogen-burning cycles: CNO, NeNa and MgAl. The effect is to convert
carbon and oxygen to nitrogen, neon to sodium and magnesium to aluminium.
The abundance anomalies seen in globular cluster stars, for example
the oxygen-sodium anticorrelation, may be due to HBB in AGB stars
\cite{Gratton}. The combination of third dredge-up of carbon with
HBB is an important source of primary nitrogen, especially at low-metallicity.

Stellar models of hot-bottom burning rely on nuclear reaction rates
that are inherently uncertain due to both experimental limits and
extrapolation into the low-energy stellar regime. Our aim is to determine
the effect of considering the rate uncertainties on the stellar abundances,
and hence the chemical yields from AGB stars undergoing HBB. We have
selected the NeNa and MgAl proton capture reactions primarily because
these are modelled directly by our synthetic AGB model. It is not
possible to vary the CNO reactions because they contribute significantly
to the stellar luminosity, so affect the stellar structure, while
we are assuming that changes in the NeNa/MgAl reaction rates do \emph{not}
affect the stellar structure.

\section{Model}

\begin{figure}
\begin{centering}\begin{tabular}{ccccccccc}
\includegraphics[bb=0bp 250bp 254bp 542bp,scale=0.08]{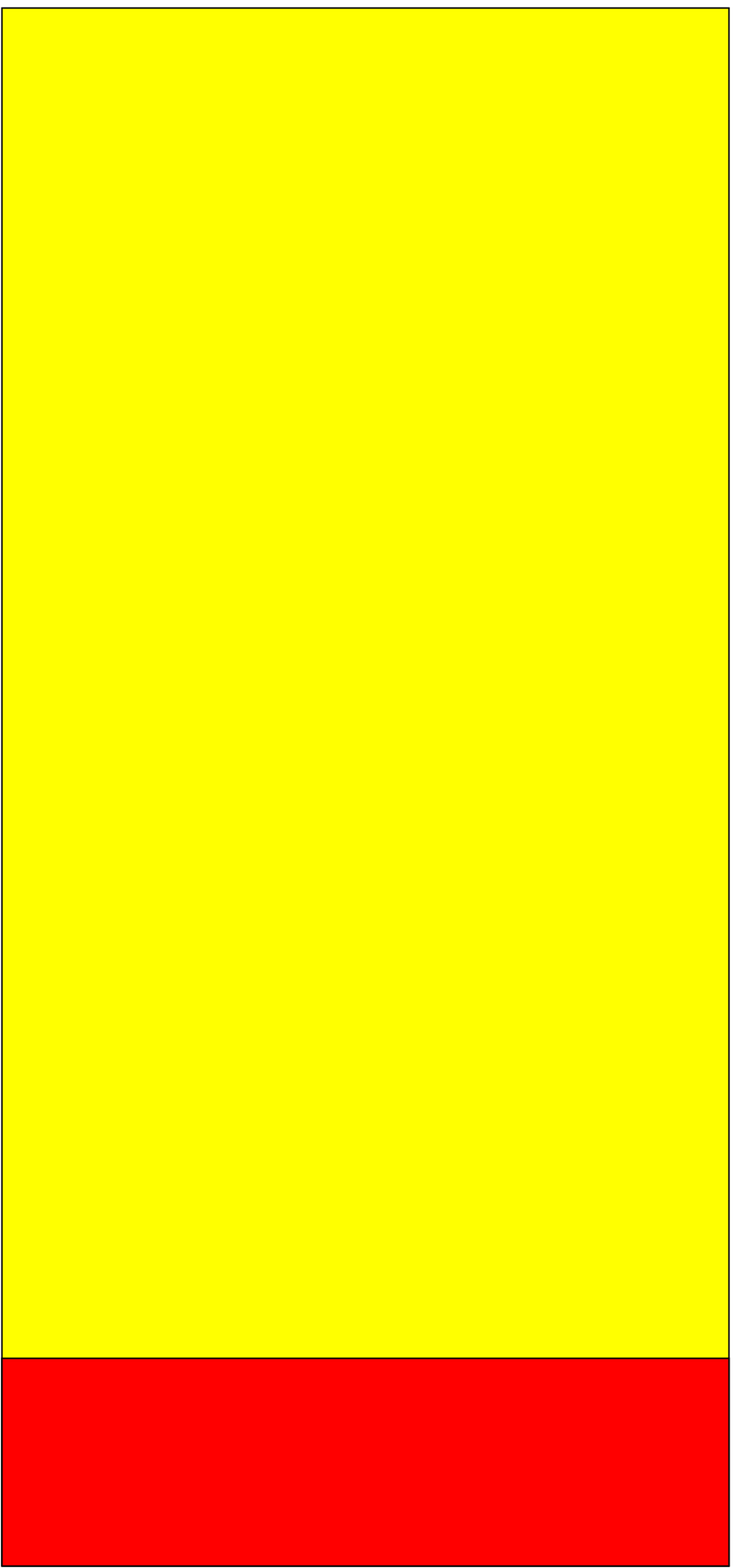}+&
\includegraphics[bb=0bp 250bp 254bp 542bp,scale=0.08]{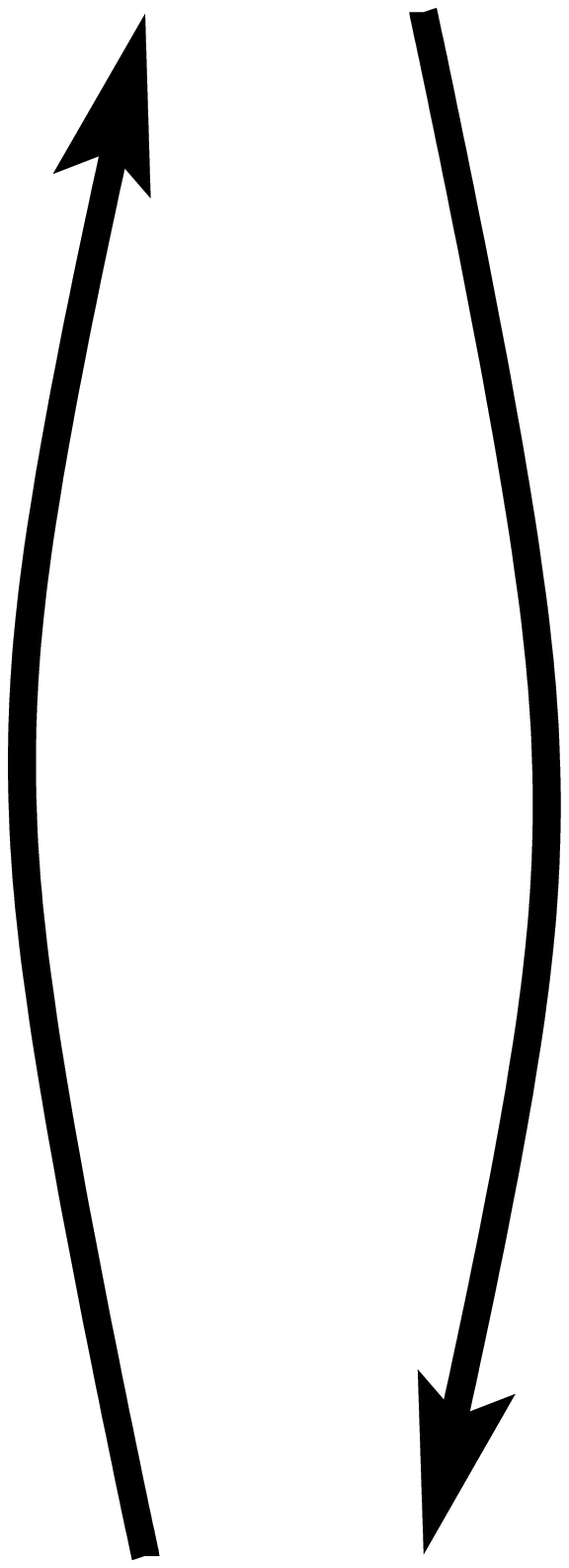}+&
\includegraphics[bb=0bp 250bp 254bp 542bp,scale=0.08]{conv_env_plain.eps}+&
\includegraphics[bb=0bp 250bp 254bp 542bp,scale=0.08]{mix.eps}+&
\includegraphics[bb=0bp 250bp 254bp 542bp,scale=0.08]{conv_env_plain.eps}+&
\includegraphics[bb=0bp 250bp 254bp 542bp,scale=0.08]{mix.eps}+&
\ldots{}&
$\approx$&
$N\times$\includegraphics[bb=0bp 250bp 254bp 542bp,scale=0.08]{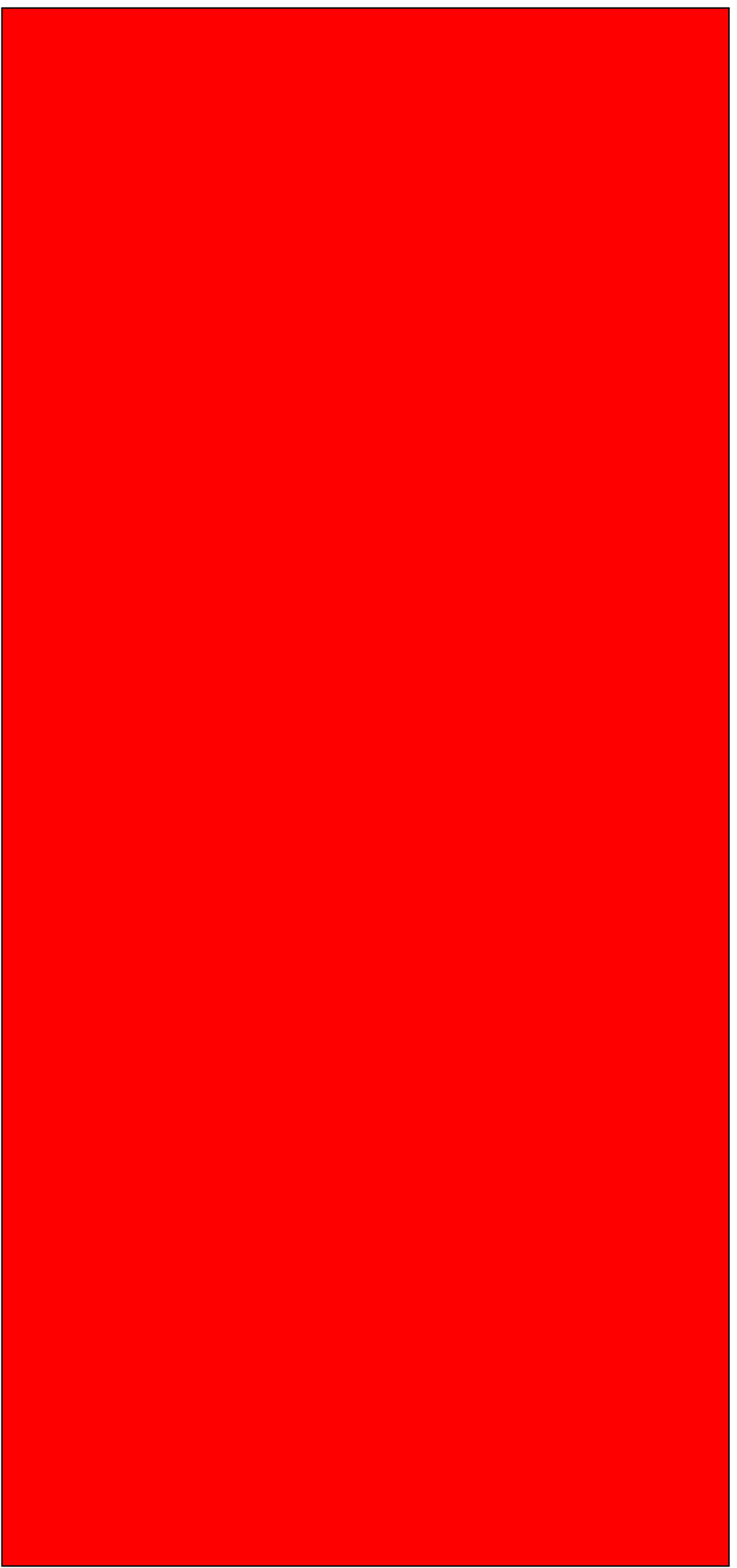}\tabularnewline
\end{tabular}\vspace{10mm}\par\end{centering}

\caption{\label{fig:Schematic-HBB}Schematic view of our hot-bottom burning
model. The detailed stellar models show that only a thin layer at
the base of the convective envelope is hot enough to burn. In reality,
and in the detailed models, the thin layer is continously mixed over
hundreds of convective turnover timescales, but our synthetic model
can reproduce the effect by burning a large fraction of the stellar
envelope for a short time.}
\end{figure}

Our synthetic AGB model is described in detail in \cite{Izzard} but
has been extended to include the NeNa and MgAl cycles (Izzard et al
2006, \emph{in preparation}) as well as the CNO cycle. We model burning
in the convective envelope by replacing the burn/mix/burn/mix\ldots{}
cycle with a single burn/mix event (see Figure \ref{fig:Schematic-HBB})
where the amount of material burnt, and the time it is burnt for,
is calibrated to detailed AGB models \cite{Karakas}. %

We analytically solve the differential equations for the abundances
of $^{20,21,22}\mbox{Ne}$, $^{23}\mbox{Na}$, $^{24,25,26}\mbox{Mg}$
and $^{26g,26m,27}\mbox{Al}$ in the burning region so we can follow
the surface abundances as a function of time. The NeNa cycle is solved
by an eigenvalue method (assuming no input to or output from the cycle,
which is justified as $^{23}\mbox{Na}(p,\gamma)^{24}\mbox{Mg}$ is
usually slow and input from $^{19}\mbox{F}$ is negligible compared
to the amount of $^{20}\mbox{Ne}$ present), then $^{23}\mbox{Na}$
is decayed explicitly into $^{24}\mbox{Mg}$. The MgAl cycle then
consists only of forward reactions if we ignore $^{27}\mbox{Al}(p,\alpha)^{24}\mbox{Mg}$,
which is always justified at the temperature of HBB ($\log_{10}T/\mbox{K}\sim7.7-8.0$),
so each rate equation is easily solved analytically. Beta-decays except
that of $^{26}\mbox{Al}^{g}$ are treated as instantaneous, we do
allow for $^{27}\mbox{Al}(p,\alpha)^{24}\mbox{Mg}$ by explicit decay
of $^{27}\mbox{Al}$ (the rate of this reaction is always negligible)
and the chain is terminated at $^{27}\mbox{Al}$ because $^{27}\mbox{Al}(p,\gamma)^{28}\mbox{Si}$
is very slow. The nuclear reactions, rate uncertainties in the HBB
temperature range and references are found in Table \ref{tab:Nuclear-reaction-rates}.%
\begin{table}
\begin{centering}{\huge }\begin{tabular}{|c|c|c|c|}
\hline 
Reaction&
Lower limit&
Upper limit&
{\tiny Reference}\tabularnewline
\hline 
$^{20}\textrm{Ne}(p,\gamma){^{21}\textrm{Na}(\beta^{+})}{^{21}\textrm{Ne}}$&
$-50\%$&
$+50\%$&
{\tiny Iliadis et al. 2001 \cite{Iliadis}=NACRE} {\tiny \cite{NACRE} }\tabularnewline
\hline 
$^{21}\textrm{Ne}(p,\gamma){^{22}\textrm{Na}(\beta^{+})}{}^{22}\textrm{Ne}$&
$-20\%$&
$+20\%$&
{\tiny Iliadis et al. 2001 \cite{Iliadis}}\tabularnewline
\hline 
$^{22}\textrm{Ne}(p,\gamma)^{23}\textrm{Na}$&
$-50\%$&
$\times2000$&
{\tiny Hale et al. 2001} {\tiny \cite{Iliadis}}\tabularnewline
\hline 
$^{23}\textrm{Na}(p,\alpha)^{20}\textrm{Ne}$&
$-30\%$&
$+30\%$&
{\tiny Rowland et al. 2004 \cite{Rowland} }\tabularnewline
\hline 
$^{23}\textrm{Na}(p,\gamma)^{24}\textrm{Mg}$&
$/40$&
$\times10$&
{\tiny Rowland et al. 2004 \cite{Rowland} }\tabularnewline
\hline 
$^{24}\textrm{Mg}(p,\gamma){^{25}\textrm{Al}(\beta^{+})}{}^{25}\textrm{Mg}$&
$-17\%$&
$+20\%$&
{\tiny Iliadis et al. 2001 \cite{Iliadis} = Powell et al. 1999 \cite{Powell} }\tabularnewline
\hline 
$^{25}\textrm{Mg}(p,\gamma){^{26}\textrm{Al}^{g}(\beta^{+})}{}^{26}\textrm{Mg}$&
$-50\%$&
$\times1.5$&
{\tiny Iliadis et al. 2001 \cite{Iliadis}}\tabularnewline
\hline 
$^{26}\textrm{Mg}(p,\gamma)^{27}\textrm{Al}$&
$/4$&
$\times10$&
{\tiny Iliadis et al. 2001\cite{Iliadis}}\tabularnewline
\hline 
$^{26}\textrm{Mg}(p,\gamma)^{27}\textrm{Al}$&
$-25\%$&
$\times3$&
{\tiny Iliadis et al. 2001\cite{Iliadis}}\tabularnewline
\hline 
$^{26}\textrm{Al}^{g}(p,\gamma)^{27}\textrm{Si}(\beta^{+})^{27}\mbox{Al}$&
$-50\%$&
$\times600$&
{\tiny Iliadis et al. 2001\cite{Iliadis}}\tabularnewline
\hline
\end{tabular}\par\end{centering}{\huge \par}

\caption{\label{tab:Nuclear-reaction-rates}Nuclear reaction rates in our
network and their uncertainties.}
\end{table}

We have constructed two sets of models in which:

\begin{enumerate}
\item Each reaction is varied individually,
\item All reactions are varied at the same time.
\end{enumerate}
When comparing the yield of an isotope, which we define simply as
the mass ejected as that isotope, we use a control model where all
the reaction rates are set to their recommended values (i.e. their
rate multipliers are all $1.0$). We use $M=5,6\mathrm{\, M_{\odot}}$
and $Z=0.02,0.004$ to test the effect of increasing mass and decreasing
metallicity, both of which increase the amount of hot-bottom burning.

\section{Results}

\begin{figure}
\includegraphics[scale=0.6,angle=270,bb = 0 0 600 100, type=eps]{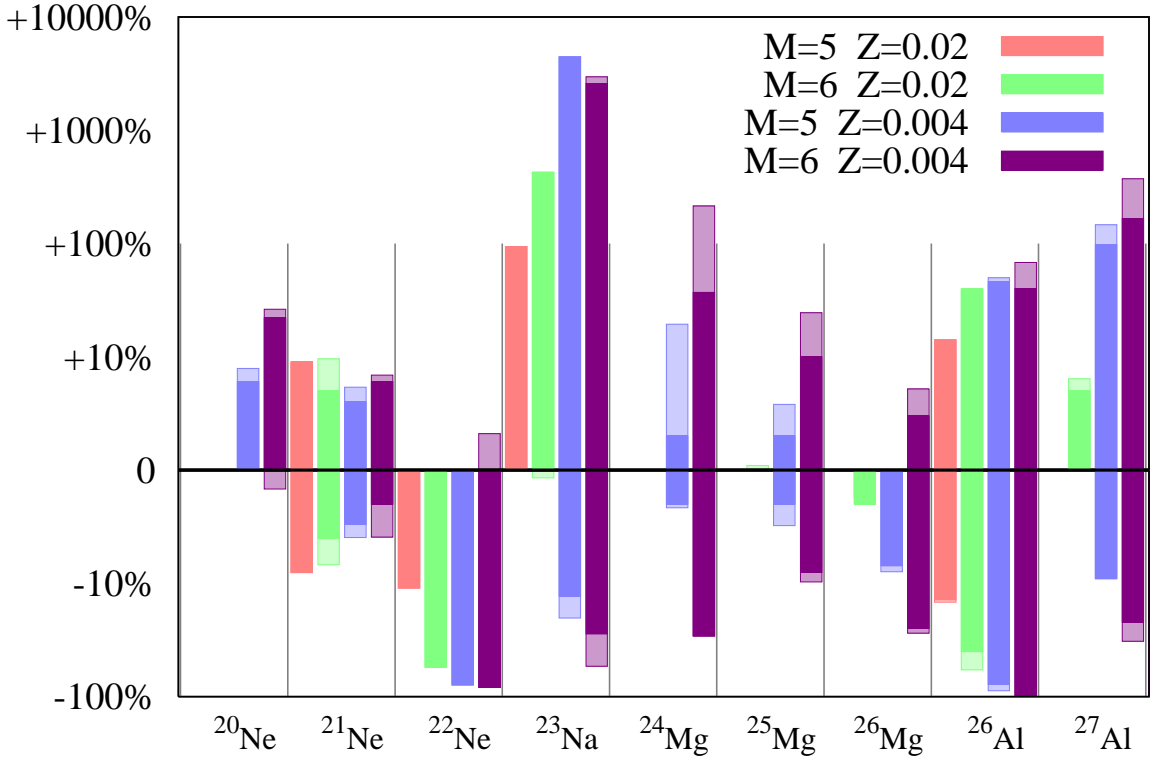}

\caption{\label{fig:Results-yields}Uncertainty ranges for chemical yields
as the difference from the control value as a function of isotope
for our four sets of model.  Dark colours are the maximum uncertainty
ranges for varying each reaction individually, the light colours give
the uncertainty when all the reactions are varied together.}
\end{figure}
Figure \ref{fig:Results-yields} shows the effect on chemical yields
of the NeNa and MgAl isotopes of varying the proton-capture nuclear
reaction rates. The values shown are percentage differences from the
control values, where positive values indicate extra production, negative
values extra destruction, and zero means no change. 

The main effects are:

\begin{itemize}
\item Complete conversion of $^{22}\mbox{Ne}$ to $^{23}\mbox{Na}$, or
not, due to the large uncertainty in the $^{22}\mbox{Ne}(p,\gamma)^{23}\mbox{Na}$
rate. The yield of $^{23}\mbox{Na}$ can be boosted by a factor of
about $40$ at $Z=0.004$, in both the $5$ and $6\mathrm{\, M_{\odot}}$
models. This sodium is \emph{primary};
\item A very large uncertainty in $^{26}\mbox{Al}$ production, ranging
from $\times0$ to $\times2$. This stems from the $\times600$ range
of the $^{26}\mbox{Al}(p,\gamma)^{27}\mbox{Si}$ rate. There is a
corresponding knock-on effect on $^{27}\mbox{Al}$;
\item Most other isotopes vary by at most about $10-20\%$.
\end{itemize}
It is sometimes necessary to consider changing all the reaction rates
together. The yields of the magnesium isotopes at $Z=0.004$ show
this clearly, as they are affected by the preceding NeNa cycle rate
changes followed by $^{23}\mbox{Na}(p,\gamma)^{24}\mbox{Mg}\dots(p,\gamma)^{25}\mbox{Al}(\beta^{+})^{25}\mbox{Mg}\dots(p,\gamma)^{26}\mbox{Al}(\beta^{+})^{26}\mbox{Mg}$.

\section{Conclusions}

We have showed that changes in yields from hot-bottom burning stars
due to nuclear reaction rate uncertainties can be significant, particularly
for $^{22}\mbox{Ne}$, $^{23}\mbox{Na}$, the magnesium isotopes and
aluminium. The effect of reaction rate uncertainty becomes stronger
as the hydrogen burning temperature increases i.e. at higher mass
and/or lower metallicity. For some isotopes, especially those of magnesium,
it is important to consider changing \emph{all} the uncertain reaction
rates at the same time, as simply changing the proton capture rate
on each isotope does not give an accurate error range for the yield.

It remains to be seen if, say, a higher $^{22}\mbox{Ne}(p,\gamma)^{23}\mbox{Na}$
rate can help to solve the problem of the globular cluster abundance
anomalies, which show excessive sodium and lower oxygen, particularly
as we have not tried to change the CNO cycle rates here. It is also
possible that the yields of primary sodium could be constrained by
galactic chemical evolution models, which in turn would give us an
upper limit for the $^{22}\mbox{Ne}(p,\gamma)^{23}\mbox{Na}$ rate,
although we will also have to examine carefully the role of massive
stars in $^{23}\mbox{Na}$ production.

Factor of two changes in the magnesium yields in our low-metallicity
models may be important for fine-structure constant ($\alpha$) variation
studies because $\Delta\alpha/\alpha$ is a steep function of $(^{25}\mbox{Mg}+^{26}\mbox{Mg})/^{24}\mbox{Mg}$
(see Fig. 1 of \cite{Fenner}).

\end{document}